\documentclass[12pt,a4paper]{article}
\pdfoutput=1
\usepackage{a4wide}
\usepackage[centertags]{amsmath}
\usepackage{amssymb}
\usepackage{cite}
\usepackage{ulem}
\usepackage{float}
\usepackage[pdftex]{graphicx,color} 
\allowdisplaybreaks
\begin{document}

\thispagestyle{empty}

\begin{center}
{\Large  \bf Tuning the Cosmological Constant, Broken Scale
  Invariance, Unitarity} 
\end{center}

\vspace*{1cm}

\centerline{Stefan F{\"o}rste and Paul Manz}
\vspace{1cm}

\begin{center}{\it
Bethe Center for Theoretical Physics\\
{\footnotesize and}\\
Physikalisches Institut der Universit\"at Bonn,\\
Nussallee 12, 53115 Bonn, Germany}
\end{center}

\vspace*{1cm}

\centerline{\bf Abstract}
\vskip .3cm
We study gravity coupled to a cosmological constant and a scale but
not conformally invariant sector. In Minkowski vacuum, scale
invariance is spontaneously broken. We consider small fluctuations
around the Minkowski vacuum. At the linearised level we find that the
trace of metric perturbations receives a positive or negative mass squared
contribution. 
However, only for the Fierz-Pauli combination the theory is
free of ghosts. The mass term for the trace of metric perturbations
can be cancelled by explicitly breaking scale invariance. This
reintroduces fine-tuning. Models based on four form field strength
show similarities with explicit scale symmetry breaking due to
quantisation conditions.  

\vskip .3cm

\newpage

\section{Introduction}

The cosmological constant problem can be looked at from various
angles, a classic and some more recent reviews are listed in \cite{Weinberg:1988cp,Nobbenhuis:2004wn,Copeland:2006wr,Polchinski:2006gy,Bousso:2007gp,Burgess:2013ara,Padilla:2015aaa}. It is
essentially the fact that known contributions to the cosmological
constant are many orders of magnitude above its observed value. In the
present note, we will discuss the question whether it is possible to
design a sector with an adjustable contribution to the cosmological
constant. We impose such a sector to be scale invariant. The different
classical solutions are related by scale transformations. Choosing one
solution amounts to tuning the cosmological constant to a given
value. On the other hand, this tuning sector should not be conformally
invariant because a traceless energy momentum tensor cannot change an
effective cosmological constant. There is a theorem that such theories
are not unitary
\cite{Luty:2012ww,Fortin:2012hn,Dymarsky:2013pqa,Dymarsky:2014zja}. However,
we  also need to mimic a cosmological constant implying that the energy
momentum tensor of the tuning sector is covariantly constant. We can
achieve that with a somewhat pathological setup for which we are not
sure whether the theorem applies. For instance, perturbations around a
classical solution can be gauged away leaving only gravity at the
linearised level.    

First, we consider a rather contrived
model capable of contributing an arbitrary cosmological constant at
the classical level. We also investigate perturbations of these classical
solutions and impose consistency conditions. In particular, gravity
should be quantisable as an effective theory. This is necessary to
maintain a quantum picture for mass attraction due to the
exchange of gravitons. We find a mass term just for the trace of
metric perturbations. In such a theory there are negative norm states
and the mass term has to be cancelled. This can be done by explicitly
breaking scale invariance. Then the observed value of the cosmological
constant is fixed in terms of its bare value and model parameters, the
tuning feature is lost. 

We suspect that the appearance of a mass term just for the trace of
metric perturbations is generically related to spontaneously broken
scale invariance. As another example we consider a tuning sector made
of four form field strength. Indeed, the same problem arises. Here,
however, scale invariance is broken by the quantisation condition
leaving the cosmological constant tunable by discrete amounts which is
crucial for the string landscape approach to the cosmological
constant problem.

\section{Minkowski Vacua}
\subsection{Classical Solution}
Our starting point is the following action
\begin{equation}
S = \int d^4x \sqrt{-g}\left\{ \frac{R}{2\kappa^2} -\Lambda
  -\lambda\left| \frac{\gamma}{g}\right|^\alpha\right\} ,
\label{eq:action}
\end{equation}
where $g$ denotes the determinant of the metric tensor $g_{\mu\nu}$,
$R$ is the scalar 
curvature and the cosmological constant $\Lambda$ is understood to
include a bare contribution together with all quantum corrections. The
tuning sector consists of four scalars $\phi^M, M \in
\left\{0,1,2,3\right\}$ which are combined into a four by four matrix
\begin{equation}
\gamma_{\mu\nu} = \partial_{\mu} \phi^M \partial_\nu \phi^N \eta_{MN} .
\label{eq:gammadef}
\end{equation}
We are using the mostly plus convention in which the Minkowski metric
$\eta_{MN} = \text{diag}\left( -1,\right.\allowbreak 1,\allowbreak
  1,\allowbreak \left. 1\right) $.  
The last term in (\ref{eq:action}) will be our tuning sector. It
is invariant under diffeomorphisms and scale transformation (to be
discussed further in section \ref{sec:conf}). The scalars $\phi^M$
could be viewed as a 
coordinate of a 
space filling brane wrapping a parallel universe. The metric induced
on the worldvolume of that brane is $\gamma_{\mu\nu}$. Parallel universes
in context of the cosmological constant have been considered e.g.\ in
\cite{Gabadadze:2014rwa} to improve an old proposal
by Tseytlin \cite{Tseytlin:1990hn}. In our case the parallel universe
is just, barring the space filling brane, empty Minkowski space. There
are no gravitational interactions in the parallel universe.
The Lorentz isometry of the parallel universe appears as global
symmetry in our universe. In a theory with quantised gravity global
symmetries are believed to be absent, see e.g.\ \cite{Banks:2010zn}.
So, finally the Lorentz isometry should be gauged. 
For now, we are not quantising gravity and neglect corresponding effects.
The last term in (\ref{eq:action}) is some weighted geometric mean between the
squareroots of the
determinants of the induced metric on the brane and our spacetime
metric. For $\alpha = 1/4$ we have the usual geometric mean, for
$\alpha =0$ just another cosmological constant, and for $\alpha = 1/2$
our space filling brane completely decouples from gravity. At the
moment, the only motivation for the last term in (\ref{eq:action}) is
that it seems to help with the fine tuning problem of the cosmological
constant.    

Variation with
respect to the metric yields Einstein's equations
\begin{equation}
R_{\mu\nu} - \frac{1}{2} g_{\mu\nu} R = - \kappa^2 \left(\Lambda 
+\lambda\left( 1 - 2\alpha\right)\left| \frac{\gamma}{g}\right|
^\alpha\right) g_{\mu\nu} .
\label{eq:einst}
\end{equation}
The field equations for the scalars $\phi^M$ read
\begin{equation}
\partial_\mu\left(\left(\sqrt{-g}\right)^{1-2\alpha} \left(
  -\gamma\right)^\alpha 
  \left(\gamma^{-1}\right) ^{\mu\nu} \partial_\nu \phi^M\right) = 0
\label{eq:scalars}
\end{equation}
where $\left(\gamma^{-1}\right) ^{\mu\nu}$ is just the $\mu\nu$
component of the inverse of $\gamma_{\mu\nu}$. i.e.\ the metric is not
involved in raising the indices. A very simple solution is
\begin{equation}
g_{\mu\nu}=\eta_{\mu\nu}\,\,\, \text{and}\,\,\, \partial_\mu \phi^M =
C\delta^M_\mu , 
\label{eq:phisol}
\end{equation}
where the integration constant $C$ is fixed by (\ref{eq:einst})  
\begin{equation}
\Lambda =   \lambda \left( 2\alpha -1\right) \left|
  C\right|^{8\alpha}.
\label{eq:intfix}
\end{equation}
So, any cosmological constant $\Lambda$ whose sign matches the sign of
the RHS of (\ref{eq:intfix}) can be sequestered from gravity in that
it does not result in spacetime curvature. Notice that
(\ref{eq:phisol}) relates the tuning sector to the vierbein of
Minkowski spacetime. This bears some similarity to the ``vacuum
variable'' of  \cite{Klinkhamer:2007pe}. Notice however, that the
tuning (\ref{eq:intfix}) does not solve the fine tuning problem of the
cosmological constant by itself. In the context of unimodular gravity
\cite{Anderson:1971pn,vanderBij:1981ym} this is discussed in
\cite{Nobbenhuis:2004wn}. Unimodular gravity is just a gauge fixed
version of Einstein gravity. Unimodularity can be imposed with a
Langrange multiplier. The Bianchi identity forces the Lagrange
multiplier to take the role of a cosmological constant which now
appears as an integration constant. Our setup is different in that we
have added a scale invariant sector. Still, as we will see later, it
does not provide a solution to the fine tuning problem of the
cosmological constant.

\subsection{Scale Invariance without Conformal
  Invariance \label{sec:conf}} 

Conformal symmetry and its relation to the cosmological constant has
been already addressed in \cite{Weinberg:1988cp} and references
therein. A more recent discussion concerning a proposal of Kaloper
and Padilla \cite{Kaloper:2013zca} can be found in
\cite{Ben-Dayan:2015nva}. Here, we will argue on a classical level
that the tuning sector in (\ref{eq:action}) is scale invariant
but not conformally invariant. Let us specify scaling
transformations in our action (\ref{eq:action}). We focus on the last
term and, for simplicity, take $g_{\mu\nu} = \eta_{\mu\nu}$, i.e.\ we
consider the action
\begin{equation}
S_{\gamma} = - \lambda \int d^4 x \left| \gamma\right| ^\alpha .
\label{eq:scale}
\end{equation}
This action is invariant under global scale transformations
\begin{equation}
x^\mu \longrightarrow \Omega\, x^\mu ,
\end{equation}
provided the four scalars transform as
\begin{equation}
\phi^M\left( x \right) = \Omega^\Delta \tilde{\phi}^M \left( \Omega
  x\right) \,\,\, \text{with} \,\,\, \Delta = \frac{1}{2\alpha} - 1.
\label{eq:phiscaling}
\end{equation}
The corresponding Noether current is
\begin{equation}
D_\mu = x^\rho T_{\mu\rho} - J_\mu .
\label{eq:scalecurr}
\end{equation}
Here, $T_{\mu\nu}$ is the energy momentum tensor which can be either
obtained by taking the variational derivative with respect to the
metric $g_{\mu\nu}$ before restricting to the Minkowski metric or by
computing the Noether current belonging to constant shift symmetry of
the coordinates $x^\mu$ in (\ref{eq:scale}). Either way it is (cf
(\ref{eq:einst})) 
\begin{equation}
T_{\mu\nu} = \lambda\left( 1 - 2 \alpha\right) \left|
  \gamma\right|^\alpha \eta_{\mu\nu}.
\end{equation}
The second term in (\ref{eq:scalecurr}) is called virial current
\cite{Nakayama:2013is}. It reflects the non vanishing scaling
dimension of our fields and is given by
\begin{equation}
J^\mu = \lambda \left( 1 -2\alpha\right) \left| \gamma\right|^\alpha
\left(\gamma^{-1}\right)^{\mu\nu} \phi^M\partial_\nu \phi^N \eta_{MN} .
\end{equation}
Using equations of motion one can check that the trace of the
energy momentum tensor is indeed given by the divergence of the virial
current, 
\begin{equation}
T^\mu_\mu = \partial_\mu J^\mu = 4\lambda \left( 1 -2\alpha\right)
\left|\gamma\right|^\alpha .
\label{eq:emtrace}
\end{equation}
The non vanishing trace is a signal that scale symmetry is strictly
global and not enhanced to conformal symmetry. This is essential,
because otherwise the tuning sector could not cancel the
cosmological constant. There could, however, be an improved energy
momentum tensor whose trace vanishes. Indeed, our trace can be
expressed as\footnote{Notice, however, $J_\mu \not= \partial_\mu L =
  D_\mu + J_\mu$, i.e.\ with our $L$ one cannot construct a
  conserved conformal current in contrast to the $L$ discussed e.g.\
  in \cite{Dymarsky:2013pqa}.} 
\begin{equation}
T_{\mu}^{\mu} = \partial^\kappa\partial^\lambda L_{\kappa\lambda}
=\Box L ,\,\,\,
\text{with}\,\,\, L_{\kappa\lambda} = \eta_{\kappa\lambda}L,\,\,\, L =
\frac{\lambda}{2}\left( 1 -
  2\alpha\right) \left| \gamma\right|^\alpha x^2,
\end{equation}
where we used that (\ref{eq:scalars}) implies constant $\gamma$
for
$\alpha \not= \frac{1}{2}$. (This can be seen by multiplying
(\ref{eq:scalars}) with $\partial_\lambda \phi _M$, summing over $M$,
using Leibniz rule and 
$\partial_\mu \gamma = \gamma \left( \gamma^{-1}\right)
^{\rho\kappa}\partial_\mu \gamma_{\rho\kappa} $.)
Then there is a traceless, conserved improved tensor
\cite{Nakayama:2013is} 
\begin{equation}
T_{\mu\nu}
+\frac{1}{2}\left( \partial_\mu \partial_\rho {L^\rho}_\nu
  + \partial_\nu \partial_\rho {L^\rho}_\mu  - \partial^2 L_{\mu\nu} -
  \eta_{\mu\nu} \partial_\rho\partial_\kappa L^{\rho\kappa}\right) +
\frac{1}{6} \left( \eta_{\mu\nu} \partial^2 {L^\rho}_\rho
  - \partial_\mu\partial_\nu {L^\rho}_\rho \right),
\end{equation}
which in our case, however, vanishes. 
A similar discussion applies to
models involving a four form field strength
\cite{Aurilia:1980xj,Duff:1980qv,Hawking:1984hk,Bousso:2000xa}. For
other, related models see e.g.\
\cite{Fukuyama:1983hv,Guendelman:1999qt,Guendelman:2013ke}. Having a 
theory which is scale but not conformally invariant indicates 
potential problems\footnote{For $\alpha = 1/2$, action (\ref{eq:scale}) is
  invariant under general coordinate transformations, i.e.\ in
  particualr under conformal transformations. 
An interesting limit may be given by  $\alpha \to \frac{1}{2}$,
  $\lambda \to \infty$ 
  such that $\left( \alpha - 1/2\right) \lambda$ remains finite. Eq.\
  (\ref{eq:einst}) shows that there is still a non vanishing
  contribution to the cosmological constant. On the other hand, as long as
  $\alpha$ differs only very little from $1/2$ conformal symmetry is
  broken and the non vanishing trace in (\ref{eq:emtrace}) suggests
  that this is also true in the limit.}. Indeed there is a theorem
stating that, provided 
Poincar{\'e} invariance holds, such a theory is not unitary
\cite{Luty:2012ww,Fortin:2012hn,Dymarsky:2013pqa,Dymarsky:2014zja}). However,
our tuning sector is a pathological theory and we are a priori not sure
whether the theorem applies. We have seen that the trace of the energy
momentum tensor can be expressed in terms of a D'Alambertian. Often
this would imply conformal invariance whereas our Lagrangian is not
invariant under local scale transformations. We will see in the next
subsection that all perturbations can be gauged away and the resulting
linearised gravity is not unitary. Hence, on that note, the theorem is
confirmed.  

\subsection{Background Field Expansion and Ghosts}

To investigate the question of unitarity we check whether our model
suffers from ghosts, at a perturbative level.  
We will consider an action obtained by expanding 
(\ref{eq:action}) around the Minkowski solution till second order in
the fluctuations. Non vanishing $\Lambda$ is assumed. We perturb metric
and scalars by small fluctuations $h_{\mu\nu}$ and $\epsilon^M$
\begin{equation}
g_{\mu\nu} = \eta_{\mu\nu} + h_{\mu\nu}\,\,\, ,\,\,\, \partial_\mu
\phi^M = C\left( \delta_\mu ^M + \partial_\mu \epsilon^M\right) .
\label{eq:pert}
\end{equation}
Expanding the Einstein-Hilbert term leads to the standard result
\begin{equation}
\int d^4x \sqrt{-g}R \approx \int d^4 x\left\{ \frac{1}{4} \partial_\lambda
  h^\rho _\rho \partial^\lambda h^\kappa _\kappa -
  \frac{1}{4} \partial_\lambda h^\rho _\kappa \partial^\lambda
  h^\kappa _\rho +\frac{1}{2}\partial_\lambda h^\lambda
  _\rho \partial^\kappa h^\rho _\kappa -\frac{1}{2}\partial_\kappa
  h^\lambda _\lambda \partial^\rho h^\kappa _\rho\right\} ,
\label{eq:linein}
\end{equation}
where indices are raised and lowered with the Minkowski metric. For
us more interesting is the expansion of the remaining two terms in
(\ref{eq:action}) resulting in
\begin{equation}
\int d^4x \sqrt{\left|g\right|} \left\{\Lambda +\lambda\left|
    \frac{\gamma}{g}\right|^\alpha\right\} \approx
-\frac{\Lambda}{1-2\alpha} \int d^4x \left\{ 2\alpha
  +\frac{1}{2}\alpha \left( \alpha -\frac{1}{2}\right) \left(
    2 \partial_\mu \epsilon^\mu - h^\lambda _\lambda\right)^2\right\} ,
\label{eq:secord}
\end{equation}
where (\ref{eq:intfix}) has been employed.
The first contribution is an irrelevant constant which does not couple
to the metric. At first, it seems that as long as the coefficient in
front of the second term is positive we have a stable solution. That
is, the Euclidean version of the action is positive definite or,
phrasing it as the authors of \cite{Klinkhamer:2007pe}, we have
positive vacuum compressibility. The condition is explicitly
\begin{equation}
\alpha \Lambda > 0 .
\label{eq:stabcon}
\end{equation}
So, even if we were allowed to introduce Lagrangians like
(\ref{eq:action}) by hand we would have to decide on the sign of the
cosmological constant we want to cancel. 

The situation is actually worse. Our spacetime is not Euclidean and we
have to worry about negative norm states. In our model, two Lorentz
groups  appear.  There is the global Lorentz symmetry rotating the
four scalars $\phi^M$ and then there is the isometry of our
solution. Our background breaks the two Lorentz groups into a diagonal
one. The scalars $\phi^M$ `metamorphose' into four-vectors
$\epsilon^\mu$. As is well known from quantum electrodynamics one
needs gauge symmetry to decouple ghosts from the space of physical
states. Here, we do not have the $U(1)$ of electrodynamics but we do
have spacetime diffeomorphisms 
\begin{equation}
x^\mu \to x^\mu + \xi^{\mu}\left( x\right) .
\end{equation}
Viewing $\xi$ as a quantity of the same order as our perturbations
(\ref{eq:pert}) the leading order transformations are
\begin{equation}
h_{\mu\nu} \to h_{\mu\nu} + \partial_\mu \xi_\nu + \partial_\nu \xi_\mu
\,\,\, ,\,\,\, \epsilon_\mu \to \epsilon_\mu + \xi_{\mu} ,
\end{equation}
and we recognize a gauge invariant combination in
(\ref{eq:secord}). Now, we could impose a (partial) gauge fixing
condition 
\begin{equation}
\partial_\mu \epsilon^\mu = 0 .
\label{eq:gaugefix}
\end{equation}
We end up with a theory of massive gravity with a mass term only for
the trace of metric perturbations. Probably this is related to
spontaneous breaking of scale invariance. It is proven in
\cite{VanNieuwenhuizen:1973fi} that, at the linearised
level\footnote{Here, `linearised' refers to the equations of motion and
  corresponds to an action quadratic in perturbations. The remaining
  orders in perturbation theory are discussed in
  \cite{deRham:2010ik,Hassan:2011hr}. For
  reviews see e.g.\ \cite{Hinterbichler:2011tt,deRham:2014zqa}.}, the
only consistent pure gravity theories are either the Einstein
Lagrangian (\ref{eq:linein}) or the Fierz-Pauli Lagrangian. The mass
term in that Lagrangian contains the combination $h_{\mu\nu}h^{\mu\nu}
-\left( h_\lambda ^\lambda\right)^2$ which does
not appear in our expansion even for particular values of
parameters. So, the only way our model can be consistent is to replace
the inequality (\ref{eq:stabcon}) by an equality. But then the model
is useless as far as the fine tuning problem for the cosmological
constant is concerned.

\subsection{Explicitly Breaking Scale Invariance}

We break scaling invariance by supplementing (\ref{eq:action}) with
another contribution
\begin{equation}
S = \int d^4x \sqrt{-g}\left\{ \frac{R}{2\kappa^2} -\Lambda
  -\lambda_1\left| \frac{\gamma}{g}\right|^\alpha - \lambda_2\left|
    \frac{\gamma}{g}\right|^\beta \right\} ,
\label{eq:breakaction}
\end{equation}
with $\beta\not= \alpha$. Assigning our previous scaling dimension
(\ref{eq:phiscaling})  to $\phi^M$ we see that the last term breaks
scaling invariance explicitly. (Actually, viewing $\lambda_2$ as an
expectation value of a field with dimension $\Delta_{\lambda_2} =
4( 1 - \beta/\alpha )$, the action (\ref{eq:breakaction}) could 
still be traced back to spontaneous breakdown of scale invariance.)
Again, we make the ansatz
\begin{equation}
g_{\mu\nu} = \eta_{\mu\nu} \,\,\, ,\,\,\, \partial_\mu \phi^M = C
\delta_\mu ^M .
\end{equation}
Einstein's equations are solved if $C$ is chosen such that
\begin{equation}
0 = \Lambda + \lambda_1\left( 1 - 2\alpha\right)
\left|C\right|^{8\alpha} + \lambda_2 \left( 1 - 2\beta\right) \left|
  C\right|^{8\beta} .
\label{eq:cosmcanc}
\end{equation}
In the linearised theory (\ref{eq:secord}) is replaced by
\begin{align}
\int d^4x \sqrt{\left|g\right|} \left\{\Lambda +\lambda_1\left|
    \frac{\gamma}{g}\right|^\alpha + \lambda_2\left|
    \frac{\gamma}{g}\right|^\beta \right\} \approx &
\int d^4x \left\{ 2\alpha\lambda_1 \left|C\right|^{8\alpha} +
  2\beta\lambda_2\left|  C\right|^{8\beta} +\right. \nonumber \\
  & \left. \hspace*{-0.6in}-\frac{1}{4}\left( \lambda_1\alpha\left( 1- 2\alpha\right)
      \left| C\right|^{8\alpha} 
    +\lambda_2\beta \left(1 - 2\beta\right) \left| C\right|^{8\beta}\right)
    \left(h^\lambda _\lambda\right)^2\right\} ,
\label{eq:secordex}
\end{align} 
where the gauge fixing condition (\ref{eq:gaugefix}) has been
imposed. Still, we do not have a chance to obtain the Fierz-Pauli
action. But we can obtain linearised Einstein by choosing
\begin{equation}
\lambda_2 =  -\lambda_1 \frac{\alpha\left( 2 \alpha
    -1\right)}{\beta\left( 2 \beta -1\right)}\left| C\right|^{8\left(
    \alpha - \beta\right)} .
\label{eq:masscanc}
\end{equation}
Equations (\ref{eq:cosmcanc}) and (\ref{eq:masscanc}) are two
conditions for one integration constant $C$ leaving us with one
fine-tuning condition. We could get another integration constant by
introducing a second set of scalars $\tilde{\phi}^M$, but this would
leave us with a massless four-vector in the linearised theory. We can
set only one four-vector to zero by gauge fixing and the remaining one
would miss the usual $U(1)$ gauge symmetry. 

An alternative way to explicitly breake scale invariance of the
tuning sector is to introduce a cutoff $M$ which should be related to
$\Lambda$ as
\begin{equation}
\Lambda \sim M^4 .
\end{equation}
Now, one could wonder whether there is a parameter region for which
the mass of the ghost is above the cutoff scale. If so, unitarity
would be effectively restored. We estimate the mass of the ghost in
appendix \ref{ap:gm} as
\begin{equation}
m_g ^2 \sim \frac{\alpha \Lambda}{M_{\text{Planck}}^2} \sim \alpha
\frac{M^2}{M_{\text{Planck}}^2} M^2 .
\end{equation}
So, if we chose 
\begin{equation}
\left| \alpha\right| \gg \frac{M_{\text{Planck}}^2}{M^2} 
\end{equation}
and the sign such that (\ref{eq:stabcon}) holds we could decouple the
ghost. However, at the same time, we would also decouple 
$h_\mu ^\mu$. Diffeomorphism invariance would be partially broken and
effectively we would have unimodular gravity. As already mentioned,
this is a gauge fixed 
version of Einstein gravity for which the usual bare cosmological constant
appears as an integration constant. 

\section{Curved Vacua}
In this section we argue that a mismatch of the fine tuning condition
derived in the previous section leads to spacetime curvature.

\subsection{De Sitter Space: Classical Solution}

Using conformal time coordinates the de Sitter metric reads
\begin{equation}
ds^2 = \frac{\rho^2}{\tau^2}\left( -d\tau^2 + \delta_{ij} dx^i
  dx^j\right) ,
\label{eq:desitter}
\end{equation}
where $i,j \in \left\{ 1,2,3\right\}$ label spatial directions. The
quantity $\rho$ is related to the observed cosmological
constant $\Lambda_{\text{obs}}$
\begin{equation}
\rho^2 = \frac{3}{\kappa^2 \Lambda_{\text{obs}}} .
\end{equation}
Here, $\Lambda_{\text{obs}}$ is the cosmological constant one would
deduce by measuring spacetime curvature and assuming a cosmological
constant as its only source. Now, relating $\partial_\mu \phi^M$ to
the vierbein of the metric (\ref{eq:desitter}) does not solve the
$\phi^M$ equations (\ref{eq:scalars}) since de Sitter space has a spin
connection which cannot identically vanish. The solution is to choose a
vierbein of a metric with the same determinant, i.e.\footnote{In the
  same way as the authors of  \cite{Guendelman:2013ke} we could 
  alternatively modify our action such that it is invariant
  under local Lorentz rotations of the $\phi^M$. Here, we are not pursuing
  this possibility further.}
\begin{equation}
\partial_\mu \phi^M = C \left\{ \begin{array}{ l l l}
    \frac{\rho^4}{\tau^4} & \text{for} & \mu = M = 0 ,\\
\delta^i _j & \text{for} & M=i,\,\,\, \mu = j , \\
0 & \text{else}. & 
\end{array}\right.
\label{eq:desitphi}
\end{equation}
Plugging this into the Einstein equation (\ref{eq:einst}) we find the
observed cosmological constant to be determined by
\begin{equation}
\Lambda_{\text{obs}} = \Lambda + \lambda \left( 1 - 2\alpha\right) \left|
  C\right|^{8\alpha} .
\label{eq:desitco}
\end{equation}
So, at this stage, it seems that by adjusting the integration constant
$C$ we can get any value for the observed cosmological constant.

\subsection{Background Field Expansion}

Here, we consider again small fluctuations around the classical
solution
\begin{equation}
g_{\mu\nu} = \overline{g}_{\mu\nu} + h_{\mu\nu} \,\,\, ,\,\,\,
\partial_\mu \phi^M = \partial_\mu \overline{\phi}^M + C \partial_\mu
\epsilon^M ,
\end{equation}
where barred quantities denote the classical solution
(\ref{eq:desitter}) and (\ref{eq:desitphi}). Here, it is useful to
split the action into two terms, of which the first one gives the
standard metric fluctuations in de Sitter space
\begin{align} 
\int d^4x \sqrt{-g}\left( \frac{R}{2\kappa^2} -
  \Lambda_{\text{obs}}\right) = &
\frac{1}{2\kappa^2} \int d^4x \sqrt{-\overline{g}} \left\{
  \frac{1}{4} \nabla_\lambda 
  h^\rho _\rho \nabla^\lambda h^\kappa _\kappa -
  \frac{1}{4} \nabla_\lambda h^\rho _\kappa \nabla^\lambda
  h^\kappa _\rho \right. \nonumber \\ &
\hspace*{-0.5in}\left. +\frac{1}{2}\nabla_\lambda h^\lambda 
  _\rho \nabla^\kappa h^\rho _\kappa -\frac{1}{2}\nabla_\kappa
  h^\lambda _\lambda \nabla^\rho h^\kappa _\rho+
  \frac{\Lambda_{\text{obs}}}{2}\left(
    h^{\kappa\lambda}h_{\kappa\lambda} - \frac{1}{2} \left( h^\lambda
      _\lambda\right)^2\right) \right\}  ,
\label{eq:lineidesit}
\end{align}
where indices are raised and lowered with $\overline{g}_{\mu\nu}$ and
covariant derivatives, denoted by nabla, are defined in terms of
Christoffel symbols computed from $\overline{g}_{\mu\nu}$.
For the remaining contribution we find
\begin{align}
\int d^4 x \sqrt{-g} \left( \Lambda - \Lambda_{\text{obs}} + \lambda
  \left| \frac{\gamma}{g}\right|^\alpha\right)
 = & 2 \alpha\lambda \left| C\right| ^{8\alpha} \int d^4 x
 \frac{\rho^4}{\tau^4} \left\{ 1 +
   \right. \nonumber \\ & \left.  + \left( \alpha
     - \frac{1}{2}\right) \left( 
     \frac{\tau^4}{\rho^4} \partial_0 \epsilon^0 +  \partial_i
     \epsilon^i -\frac{1}{2} h_\lambda
     ^\lambda\right)^2\right\} .
\label{eq:quadesit}
\end{align}
The expression in the second line is again the combination invariant
under diffeomorphisms. Indeed, with
\begin{equation}
\delta h^\mu _\mu = 2 \nabla _\mu \xi^\mu\,\,\, \text{and} \,\,\,
C\delta \epsilon^M =  \xi^\mu\partial_\mu \overline{\phi}^M ,
\end{equation}
one finds
\begin{align}
\delta h^0 _0 & = 2\partial_0 \xi^0 -\frac{2}{\tau} \xi^0 , \\
\delta h^i _i & = 2 \partial_i \xi^i - \frac{6}{\tau} \xi^0 ,
  \\
\partial_0 \delta \epsilon^0 & = \frac{\rho^4}{\tau^4} \left(\partial_0
\xi^0 -\frac{4}{\tau}\xi^0\right), \\
\partial_i \delta \epsilon^i & =  \partial_i \xi^i . 
\end{align}
This symmetry can be used to remove the $\epsilon^M$ from
(\ref{eq:quadesit}) and we end up with a mass term for metric fluctuations
which is not of the Fierz-Pauli form\footnote{Massive gravity in
  curved space is e.g.\ reviewed in section 5 of
  \cite{Hinterbichler:2011tt}.}. To cancel this mass term we 
consider the modification (\ref{eq:breakaction}). Then
(\ref{eq:desitco})
is replaced by
\begin{equation}
\Lambda_{\text{obs}} = \Lambda + \lambda_1 \left( 1 - 2\alpha\right) \left|
  C\right|^{8\alpha} + \lambda_2 \left( 1 - 2\beta\right) \left|
  C\right|^{8\beta} .
\label{eq:ccpred}
\end{equation}
Imposing the last terms in (\ref{eq:lineidesit}) to be the only mass
terms for metric perturbations yields again condition
(\ref{eq:masscanc}). From that the integration constant $C$ is
determined and (\ref{eq:ccpred}) provides a prediction for the
observed value of the cosmological constant in terms of model
parameters. So far, our ansatz implied a positive constant on the
LHS of (\ref{eq:ccpred}). If parameters provide a negative prediction
we have to replace de Sitter space by Anti de Sitter space which we
briefly discuss in the next subsection.

\subsection{Anti De Sitter Space}

For anti de Sitter space the discussion is very similar. Now the
metric reads
\begin{equation}
ds^2 = \frac{\rho^2}{z^2}\left( - dt^2 + dx^2 +dy^2 + dz^2\right) ,
\end{equation}
where $\rho^2$ is related to the observed cosmological constant
according to
\begin{equation}
\rho^2 = -\frac{3}{\kappa^2 \Lambda_{\text{obs}}} .
\end{equation}
The solution for the scalars is
\begin{equation}
\partial_\mu \phi^M = C \left\{ \begin{array}{ l l l}
    \frac{\rho^4}{z^4} & \text{for} & \mu = M = 3 ,\\
\delta^M _\mu & \text{for} & M, \mu \in \left\{ 0,1,2\right\} , \\
0 & \text{else}. & 
\end{array}\right.
\end{equation}
The rest of the discussion follows quite closely the de Sitter case
with the conclusion that (\ref{eq:masscanc}) has to be imposed. 
Then (\ref{eq:ccpred}) serves as a prediction for the observed
cosmological constant, in case the RHS of (\ref{eq:ccpred}) is
negative.

\section{Relation to Four Form Field Strength Scenario}

One of the, at least naively, simplest tuning mechanisms is based
on adding a four form field strength
\cite{Aurilia:1980xj,Duff:1980qv,Hawking:1984hk,Bousso:2000xa} 
\begin{equation}
S = \int d^4 x \sqrt{-g}\left( \frac{R}{2\kappa^2} -\Lambda -
  \frac{Z}{2\cdot 4!} F_4^2\right) ,
\label{eq:ac4}
\end{equation}
where $F_4 = dA_3$ is the field strength of a three form gauge
potential and
\begin{equation}
F_4^2 = F_{\mu\nu\lambda\kappa} F^{\mu\nu\lambda\kappa} = 4!
\frac{f^2}{g}\,\,\, \text{with}\,\,\, f= F_{0123} .
\end{equation}
Let us ignore first any quantisation condition on the four form which
we will consider later. Then the $F_4$ part of our action is scale
invariant without being conformally invariant. The equations of motion are
\begin{align}
\nabla_\mu F^{\mu\nu\kappa\lambda} = & 0 ,\label{eq:3f} \\
R_{\mu\nu}- \frac{1}{2} g_{\mu\nu} R = & - \kappa^2\left( \Lambda -
  Z\frac{f^2}{2g}\right) g_{\mu\nu} \label{eq:ei4}.
\end{align}
The solution to (\ref{eq:3f}) is
\begin{equation}
f = C\sqrt{-g} ,
\end{equation}
with $C$ being an integration constant. Plugging this into the
Einstein equation (\ref{eq:ei4}) yields a maximally symmetric solution
corresponding to
\begin{equation}
\Lambda_{\text{obs}} = \Lambda + \frac{ZC^2}{2} .
\end{equation}
Now, we assume that $C$ is chosen such that $\Lambda_{\text{obs}} = 0$, and
consider perturbations
\begin{equation}
g_{\mu\nu} = \eta_{\mu\nu} + h_{\mu\nu}\,\,\, ,\,\,\, f = C \left(1+
  \epsilon\right) \,\,\, \text{with} \,\,\, C \epsilon
= \partial_{[0}\alpha_{123]} ,
\end{equation}
where $\alpha_3$ is the perturbation of the three form gauge potential.
The interesting part from the expansion is
\begin{equation}
\frac{Z}{2}\int d^4x \left(\sqrt{-g} C^2 + \frac{
    f^2}{\sqrt{-g}}\right) = \frac{ZC^2}{2}\int d^4x\left\{ 2 +\left(
  \epsilon - \frac{h^\lambda _\lambda}{2}\right)^2 \right\}.
\label{eq:seco4}
\end{equation}
The combination inside the last square is again gauge
invariant. ($F_{0123}$ transforms into $F_{0123}$ times the
determinant of the transformation matrix.) We can employ
diffeomorphisms to 
gauge away $\epsilon$ and obtain once again a mass term just for the
trace of the metric perturbation, and not for the Fierz-Pauli
combination. Probably this is a quite generic feature of spontaneously
broken scale invariance. 

Now, scale invariance is broken explicitly by a quantisation condition
on $F_4$ \cite{Bousso:2000xa}
\begin{equation}
f = \frac{e n}{Z} \sqrt{-g} .
\label{eq:quantcond}
\end{equation}
Thereby, the three form gauge field ceases to be a dynamical field and
it is natural to plug (\ref{eq:quantcond}) into the action (\ref{eq:ac4})
yielding
\begin{equation}
\Lambda_{\text{obs}} = \Lambda - \frac{e^2n^2}{2Z} 
\end{equation}
and perturbations just lead to linearised Einstein theory possibly in
a curved background. However, we could also follow \cite{Duff:1989ah}
and plug (\ref{eq:quantcond}) into the equation of motion resulting in
a sign difference
\begin{equation}
\Lambda_{\text{obs}} = \Lambda + \frac{e^2n^2}{2Z} .
\end{equation}
In this picture, the quantisation condition (\ref{eq:quantcond})
relates gauge form and metric perturbations
\begin{equation}
2 \epsilon = h^\lambda _\lambda .
\end{equation}
The unwanted mass term in (\ref{eq:seco4}) drops out for this
configuration. So, from either perspective, breaking scale invariance by
the quantisation condition cancels the mass term for the trace of
metric perturbations. The cosmological constant is predicted by the
value of the
four form. The great advantage of this scenario has been pointed
out in  \cite{Bousso:2000xa}. String theory can provide many four form
contributions. The observed value of the cosmological constant depends
on the radius of a vector in a multi dimensional charge lattice. If
the dimension is big enough it can actually cancel a Planck sized
$\Lambda$ with the required precision. For a recent philosophical
discussion see \cite{Polchinski:2016xto}.
\section{Conclusions}
The question whether it is possible to design a sector with an adjustable
contribution to an effective cosmological constant received, once
again, a negative answer. Still, we believe that 
our investigation led to some interesting theoretical insight. As a
tuning sector, we
envisaged a scale but not conformally invariant model. There is a
theorem stating that such a theory is not unitary
\cite{Luty:2012ww,Fortin:2012hn,Dymarsky:2013pqa,Dymarsky:2014zja}. Indeed,
if we decouple the tuning sector from everything else and expand around a
classical solution we find massless vectors without gauge
invariance. Such a theory is not unitary. However, the sector couples
to gravity and, via gravity, to another non scale invariant
contribution, the bare cosmological constant. Taking perturbations of
the metric into account we find a diffeomorphism invariant
action. Using that symmetry to gauge away the 
massless vectors yields massive gravity with the wrong mass term,
however. The non unitarity has been `transferred' to gravity. Our
example indicates that the theorem about non unitarity of scale but
not conformally invariant systems holds also when such systems are
coupled via gravity to a scale invariance breaking sector. In our
setup we can cancel the graviton mass term by explictly breaking scale
invariance and introducing fine-tuning. In models based on four form
field strength similar considerations apply. Here, scale invariance is
broken explicitly by a quantisation condition. This removes the
graviton mass term from the perturbed action.   
\section*{Acknowledgements}  
This work was supported by the SFB-Transregio TR33 ``The Dark
Universe'' (Deutsche Forschungsgemeinschaft) and by ``Bonn-Cologne
Graduate School for Physics and Astronomy'' (BCGS).

\begin{appendix}
\section{Estimating the Ghost Mass \label{ap:gm}}
In this appendix we give a rough estimation of the ghost mass for
linearised gravity with a mass term being off the Fierz-Pauli tuning,
that is we take the action
\begin{equation}
\int d^4x \left\{ M_{\text{Planck}}^2\sqrt{-g}R - f^4\left( a
      h_{\mu\nu}h^{\mu\nu} + b \left(h^\mu _\mu\right)^2\right)\right\} ,
\end{equation} 
where the first term is to be replaced by its linearised version
(\ref{eq:linein}). We consider dimensionless fields,
$M_{\text{Planck}}$ and $f$ have mass dimension one. The Fierz-Pauli
tuning corresponds to $a+b = 0$. For the case of small non zero $a+b$
the mass of the ghost has been estimated in
\cite{ArkaniHamed:2002sp}. In the following we extend the argument for
finite $a+b$. First, we leave the so called unitary gauge and
restore diffeomorphism invariance by replacing
\begin{equation}
h_{\mu\nu}  \to h_{\mu\nu} + 2\partial_{(\mu} \chi_{\nu)} ,
\label{eq:gaugeen}
\end{equation}
where $\chi_\mu$ transforms as vector and has mass dimension $-1$. The
ghost comes from the longitudinal part of $\chi$. Since we are just
interested in the ghost we consider
\begin{equation}
\chi_\mu = \frac{1}{f^2}\partial_{\mu} \pi ,
\end{equation}
with $\pi$ being a scalar of mass dimension zero. The mass term will
lead to mixed terms 
\begin{equation}\label{eq:mixing}
-4f^2\left( a h^{\mu\nu}\partial_\mu\partial_\nu \pi + b h^\mu _\mu \Box
  \pi\right) . 
\end{equation}
There is also a term quadratic in $\pi$ containing four
derivatives. This term signals the appearance of a ghost.  
For the Fierz-Pauli tuning ($b=-a$)  the term quadratic in $\pi$ vanishes and
the mixing of $h$ and $\pi$ in (\ref{eq:mixing}) can be cancelled by a
Weyl transformation  
\begin{equation}
h_{\mu\nu} = \hat{h}_{\mu\nu} +4\frac{a f^2}{M_{\text{Planck}}^2}
\eta_{\mu\nu} \pi .
\end{equation}
After this redefinition the action reads ($\hat{g}_{\mu\nu} =
\eta_{\mu\nu} + \hat{h}_{\mu\nu}$)
\begin{align}
& \int d^4x \left\{
  M_{\text{Planck}}^2\sqrt{-\hat{g}}\hat{R}-f^4\left( 
    a \hat{h}^{\mu\nu}\hat{h}_{\mu\nu} + b \left(\hat{h}^\mu
        _\mu\right)^2\right)\right. \nonumber\\
&\left.   -\frac{8a
      f^4}{M_{\text{Planck}}^2}\left( 5a + 8b\right) \pi \Box \pi
-4f^2\left( a + b\right) \hat{h}^\mu _\mu \Box \pi - 4\left(
  a+b\right) \Box \pi \Box \pi\right.\nonumber\\
&\left. -\frac{8af^6}{M_{\text{Planck}}^2} \left( a + 4b\right)
\hat{h}_\mu^\mu\, \pi -\frac{64a^2 f^8}{M_{\text{Planck}}^4}\left( a +
    4b\right) \pi^2 \right\}\label{eq:redef}
\end{align}
Following \cite{ArkaniHamed:2002sp}, we want to identify the ghost
mass as the momentum scale at which the kinetic energy
changes sign. For this reason, we  ignore the mass terms in the
last line of (\ref{eq:redef}). In the near Fierz-Pauli limit ($a+b
\sim 0$) the coupling between $\pi$ and metric perturbations can be
neglected and therefore the authors of \cite{ArkaniHamed:2002sp} focused
just on the $\pi$ sector (first and last term in second line of
(\ref{eq:redef})). We are ultimately interested in the $a=0$ case for
which near Fierz-Pauli and massless limit coincide. Therefore, we will
consider finite $a+b$. It is useful to reorganise metric perturbations
as follows. 
First, we
decompose them into traceless ($\bar{h}_{\mu\nu}$) and trace
($\hat{h} = \hat{h}_\rho ^\rho$)  part
\begin{equation}
\hat{h}_{\mu\nu} = \bar{h}_{\mu\nu} + \frac{1}{4}\eta_{\mu\nu}
\hat{h} .
\end{equation}
Further we transform our expressions to momentum space where we
restrict momenta to the rest frame,
\begin{equation}
p_\mu = p_0 \delta^0 _\mu .
\end{equation}
We label spatial (transverse) directions with latin indices, $i,j \in
\left\{ 1,2,3\right\}$. It is useful to split the transverse metric
variation again into traceless ($\tilde{h}_{ij}$) and trace part,
\begin{equation}
\bar{h}_{ij} = \tilde{h}_{ij} + \frac{1}{3} \delta_{ij} \bar{h}_{00} ,
\end{equation}
where we have incorporated the tracelessnes of $\bar{h}_{\mu\nu}$.
The linearised Einstein-Hilbert action (\ref{eq:linein}) gives the
following contribution to the kinetic energy
\begin{equation}
\frac{E^{\text{EH}}_{\text{kin}}}{p_0^2} = M_{\text{Planck}}^2 \left( \frac{1}{4}
  \tilde{h}_{ij}\left( p\right)^*\tilde{h}^{ij}\left( p\right) -
  \frac{3}{32}\left| \hat{h}\left( p\right) 
    -\frac{4}{3} \bar{h}^0 _0\left(p\right)\right|^2  \right).
\end{equation}
We observe that $\bar{h}^i _0$ as well as the combination $\hat{h}
+\frac{3}{4} \bar{h}^0 _0$ do not contribute. The reason is gauge
invariance of the linearised Einstein-Hilbert action. Even with the
terms coupling to $\pi$ there is an invariance under trace ($\hat{h}$)
preserving transformations. For $a\not=0$ this symmetry is broken by
the mass term (as long as we do not transform $\pi$). So at least for
$a = 0$ we can gauge fix
\begin{equation}
\bar{h}^0_0 = -\frac{4}{3} \hat{h} \,\,\, ,\,\,\, \bar{h}^0 _i =0 .
\label{eq:gfix}
\end{equation}
Fixing of symmetries involving $\pi$ transformations will be discussed
shortly. We find for the kinetic energy of the $\pi$, $\hat{h}$
directions  
\begin{align}
\frac{E_{\text{kin}}}{p_0^2}\sim  & \left( 
\hat{h}\left(p\right)^* , \pi\left(
    p\right)^*\right) \times\nonumber \\
& \hspace*{-0.1in} \left( \begin{array}{cc} 
     -\frac{625 M_{\text{Planck}}^2}{864} &  -2\left( a +
  b\right)f^2  \\ 
 -2\left( a + b\right)f^2 & 
 -\frac{8af^4}{M_{\text{Planck}}^2}\left( 5a + 8b\right)  -
   4\left(a+b\right) p_0^2 \end{array}\right) \left(\begin{array}{c}
\hat{h}\left( p\right) \\
 \pi\left( p\right) \end{array}\right) .
\label{eq:matrixdec} 
\end{align}
One of the eigenvalues of the kinetic mixing matrix will change sign at
a certain scale. The corresponding eigenvector is identified with the
ghost direction. The other direction can be removed by fixing the
(additional) gauge symmetry introduced in (\ref{eq:gaugeen}). 
Now, it is easy to find a value of $p_0^2$ at which the determinant of
the kinetic mixing matrix changes sign. Taking that value to be
the ghost mass ($m_g$) we obtain an estimate
\begin{equation}
m_g ^2 \sim \frac{6a^2f^4}{M_{\text{Planck}}^2\left( a
  + b\right)} -\frac{16 a f^4}{M_{\text{Planck}}^2} + \frac{ 864 f^4
\left( a + b\right)}{625 M_{\text{Planck}}^2} + \ldots,
\label{eq:massest}
\end{equation}
where we have organised contributions in powers of $a+b$.
For $a \not=0$ the gauge fixing in (\ref{eq:gfix}) is not
justified. What one could do instead is to include mass terms and
modify our definition of the ghost mass such that it corresponds to
the scale at which the total energy changes sign. The dots in
(\ref{eq:massest}) stand for corrections vanishing for $a =0$. 
For the case $a =0$ we see that the ghost mass is of the same order as
the $h^\mu _\mu$ mass.

\section{Combining with Extra Dimensions}
In this appendix we discuss some aspects of our setup when combined
with the idea of extra dimensions, in particular with warped brane
worlds. The best known example for such brane worlds are the
Randall-Sundrum models \cite{Randall:1999ee,Randall:1999vf}. In these
models the fine-tuning problem of the cosmological constant is
translated into matching conditions arising from delta-function
sourced five dimensional Einstein equations \cite{DeWolfe:1999cp}.
Attempts to turn this into a self-tuning
\cite{Kachru:2000hf,ArkaniHamed:2000eg} are problematic
\cite{Forste:2000ps,Csaki:2000wz}. Unconventional
Lagrangians seem to improve the situation
\cite{Kim:2000mc,Pospelov:2004aw,Antoniadis:2010ik}.
However, problems may arise when taking into account perturbations
\cite{Medved:2001ad,Forste:2011hq,Forste:2013hc} or energy conditions
\cite{Antoniadis:2014ioa}.     

In this appendix, we discuss a model with unconventional Lagrangian. It
provides a Randall-Sundrum II geometry, i.e.\ a single brane and exponential
warping in the bulk. Matching conditions just fix integration
constants. However, the effective four dimensional theory is the one
discussed in the bulk of the present paper. There, we have seen that
perturbations lead to problems. We refrain from investigating
perturbations including additional fields in the extra dimensional
scenario. We do not expect that they can improve the model.

The action is the sum of a five dimensional bulk term and a 3-brane
source
\begin{equation}
S = S_{\text{bulk}} + S_{\text{brane}} .
\end{equation}
For the bulk action we take a five dimensional version of
(\ref{eq:action})
\begin{equation}
S = \int d^4x dy \sqrt{-G}\left\{ \frac{R}{2\kappa_5 ^2} 
  -\lambda _5\left| \frac{\gamma}{G}\right|^\alpha\right\} ,
\end{equation} 
the fifth direction is called $y$. Metric and curvature are tensors in
five dimensions and $\gamma$ is defined as in (\ref{eq:gammadef}) with
the modification that $M,N \in \left\{\ 0,1,2,3,5\right\}$,
$\eta_{MN}$ is the five dimensional Minkowski metric (again in mostly
plus convention) and we used capital $G_{MN}$ for the five dimensional
metric. The brane source is described by
\begin{equation}
S_{\text{brane}} = - \left. \int d^4x \sqrt{-G} f\left( \phi^5\right)
\right|_{y=0} .
\end{equation}
(Strictly speaking, $G$ should be replaced by the determinant of the
metric induced on the brane. With ansatz (\ref{eq:warped}) there will
be no difference.) 
For $f$ being constant this is just the 4d vacuum energy containing
a classical part plus contributions due to vacuum fluctuations of all
4d particle physics fields (which are living on the brane). The
$\phi^5$ dependence can be included to allow for some non vanishing
coupling while preserving four dimensional Lorentz invariance in field
space. 

For the five dimensional metric we impose a warped ansatz
\begin{equation}
ds^2 = a\left( y\right)^2 \eta_{\mu\nu} dx^\mu dx^\nu + dy^2 ,
\label{eq:warped}
\end{equation}
where $\mu ,\nu \in \left\{0,1,2,3\right\}$. Einstein's equations provide
two independent equations (prime on $a$ denotes derivation with respect to
$y$) 
\begin{align}
6\left( \frac{a^\prime}{a}\right)^2 = & - \lambda _5 \kappa_5^2 \left( 1
  - 2\alpha\right) \left| a\right|^{-8\alpha}\left|
  \gamma\right|^\alpha , \label{eq:firste}\\
3\frac{a^{\prime\prime}}{a} + 3 \left( \frac{a^\prime}{a}\right)^2 = &
- \lambda _5 \kappa_5^2 \left( 1 
  - 2\alpha\right) \left| a\right|^{-8\alpha}\left|
  \gamma\right|^\alpha -\kappa_5^2 f\left( \phi^5\right) \delta\left(
  y\right) .\label{eq:sece}
\end{align}
The equations for the scalars read
\begin{equation}
2 \alpha\lambda _5 \partial_M\left( \left| a\right|^{4 - 8\alpha} \left|
    \gamma\right|^\alpha \left( \gamma^{-1}\right)^{MN}\partial_N \phi
  ^K\right) = \delta^K _5 f^\prime \left( \phi^5\right) \left|
  a\right|^4 \delta\left( y\right) .
\label{eq:fscalars}
\end{equation}
Our ansatz for the scalars is
\begin{equation}
\partial_M\phi^N = \left\{ \begin{array} {l l l}
C \delta_M ^N & \text{for} & M,N \in \left\{0,1,2,3\right\},\\
\varphi^\prime \left( y\right) & \text{for} & M=N=5, \\
0 & \text{else,}& \end{array}\right.
\end{equation}
where $C$ is constant. 
(Continuity of the configurations in the first line would allow to
absorb the constant in a redefinition of the first four scalars. It
will be useful to keep it, though.)
This ansatz solves (\ref{eq:fscalars})
automatically for $K \in \left\{ 
0,1,2,3\right\}$, whereas the fifth equation reads for $y\not= 0$
\begin{equation}
\partial_y \left( \left|\frac{\varphi^\prime}{a^4}\right|^{2\left(
      \alpha -1\right)} \frac{\varphi^\prime}{a^4} \right) = 0.
\end{equation}
So, for $y\not=0$ warp factor and fifth scalar are related by
\begin{equation}
\varphi^\prime = c_1 a^4 ,
\end{equation}  
where $c_1$ is another integration constant.
 Plugging this into (\ref{eq:firste}) yields
\begin{equation}
\frac{a^\prime}{a} = \mp A \left| C^4 c_1 \right|^{\alpha} ,
\label{eq:firstsol}
\end{equation}
with 
\begin{equation}
A = \sqrt{\frac{\kappa_5^2 \left( 2\alpha-1\right) \lambda _5 }{6}} ,
\label{eq:Adef}
\end{equation}
and we take the squareroot of a positive real number to be positive.
Reality of the solution implies the condition
\begin{equation}
\lambda _5\left( 2\alpha -1\right) > 0 .
\end{equation}
Inserting (\ref{eq:firstsol}) into (\ref{eq:sece}) yields no new
equation for $y \not= 0$.
Equation (\ref{eq:firstsol}) is solved by   
\begin{equation}
a = \text{exp}\left[\mp A \left| C^4 c_1 \right|^{\alpha} y\right],
\end{equation}
where for $y>0$ ($y<0$) we chose the upper (lower) sign to obtain a
finite effective Planck mass in four dimensions. Since $a$ has to be
continuous a constant factor can be absorbed into a rescaling of the
$x^\mu$. The solution
including delta function sources is now obtained by taking solutions
for $y>0$ and $y<0$ with different integration constants and
integrating equations over an infinitesimal interval containing $y=0$.  
This procedure yields so called matching conditions relating
integration constants at the two sides of the brane. The first four
scalar equations imply that $C$ should be the same on both sides of
the brane. 
Einstein equation (\ref{eq:sece}) yields a jump condition on $a^\prime$
\begin{equation}
a^\prime \left( + 0\right) - a^\prime \left( -0\right) = -\frac{1}{3}
a \kappa_5 f_{| y=0} .
\end{equation}
For integration constants this implies
\begin{equation}
A \left|C\right|^{4\alpha}\left( \left| c_1 ^>
  \right|^{\alpha}+\left| c^< _1 \right|^{\alpha}\right) =
\frac{1}{3}\kappa_5 ^2 f _{| y=0} .
\label{eq:match1}
\end{equation}
where superscript $>$ ($<$) characterises the solution for $y>0$
($y<0$).  The LHS of (\ref{eq:match1}) is never negative and we
conclude that the brane has to have positive tension (as in
Randall-Sundrum II). The fifth scalar equation (\ref{eq:fscalars})
yields another jump condition resulting in
\begin{equation}
\left|c_1 ^>\right|^{2\left( \alpha -1\right)} c_1 ^> -
\left|c_1 ^<\right|^{2\left( \alpha -1\right)}  c_1 ^< =
\frac{\left| C\right|^{-8\alpha}}{2\alpha \lambda _5} f^\prime\left(
  \phi^5\right)_{| y=0} .
\end{equation}
Notice that, up to a subtlety concerning sign, $c_1$ and $C$ always
enter in the same multiplicative combination. So, we should count them
as one integration constant. Taking into account the two sides of the
brane we still have two integration constants with two matching
conditions. So, it seems we have succeeded to obtain a Randall-Sundrum
II brane world without fine tuning. They differ, however, in the
effective four dimensional theory.

To get an idea about the effective four dimensional theory we first
freeze as many moduli as we consistently can, meaning the effective four
dimensional theory should be invariant under four dimensional
diffeomorphisms with Minkowski space being a solution. It 
turns out that we have to keep more moduli than just the four
dimensional metric as in Randall-Sundrum II. Namely, if we just
modified (\ref{eq:warped}) to
\begin{equation}
ds^2 = a\left( y\right)^2 g_{\mu\nu}\left( x\right) dx^\mu dx^\nu + dy^2 ,
\end{equation}
plugged in the solution for the rest and integrated over the fifth
direction only the Einstein-Hilbert term and the brane source would
give rise to a diffeomorphism invariant theory, 
\begin{equation}
S_1 = \int d^4x \sqrt{-g}\left\{ \frac{R}{2\kappa^2} - \Lambda\right\}
.
\end{equation}
Here, the effective gravitational coupling is
\begin{equation}
\frac{1}{\kappa ^2} = \frac{\left| C\right|^{-4\alpha}}{2A\kappa_5
  ^2}\left( \left| c_1 ^> 
    \right|^{-\alpha} +  \left| c_1 ^< \right|^{-\alpha} \right)  ,
\end{equation}
whereas 
\begin{equation}
\Lambda = \frac{1}{2} f_{| y=0} . 
\label{eq:effla}
\end{equation}
Minkowski space is not a solution which is no surprise since we have not
yet taken into account the scalars. To obtain a
diffeomorphism invariant theory we have to keep the first four scalars
$\phi^\mu \left( x\right)$ as moduli. This yields a contribution
\begin{equation}
S_2 = -\lambda \int d^4x \sqrt{-g} \left| \frac{\gamma}{g}
\right|^\alpha,
\end{equation}
where $\gamma$ is now the determinant of the four dimensional matrix
as in (\ref{eq:gammadef}) and
\begin{equation}
\lambda  = \lambda_5 \int dy \left| a^4\right|^{1-2\alpha} \left|
  \varphi^\prime\right|^{2\alpha} =
\frac{\lambda_5}{4A}\left|C\right|^{-4\alpha}\left( \left|
    c_1^>\right|^\alpha +\left| c_1 ^<\right|^\alpha\right) .
\end{equation}
Using (\ref{eq:match1}), (\ref{eq:Adef}) and (\ref{eq:effla}) this can
be rewritten into (\ref{eq:intfix}). Although we obtained the
Randall-Sundrum geometry the effective four dimensional theory is
different. Barring the frozen moduli, it is the same model as
discussed in the bulk of the paper. We have seen, linearised gravity
cannot be consistently quantised in that model. We do not expect 
taking into account the frozen moduli will provide an improvement.    
\end{appendix}


\begin{thebibliography}{99}
%
\bibitem{Weinberg:1988cp}
  S.~Weinberg,
  Rev.\ Mod.\ Phys.\  {\bf 61} (1989) 1.
%
\bibitem{Nobbenhuis:2004wn}
  S.~Nobbenhuis,
  Found.\ Phys.\  {\bf 36} (2006) 613
  [gr-qc/0411093].
%
\bibitem{Copeland:2006wr}
  E.~J.~Copeland, M.~Sami and S.~Tsujikawa,
  Int.\ J.\ Mod.\ Phys.\ D {\bf 15} (2006) 1753
  [hep-th/0603057].
%
\bibitem{Polchinski:2006gy}
  J.~Polchinski,
  hep-th/0603249.
%
\bibitem{Bousso:2007gp}
  R.~Bousso,
  Gen.\ Rel.\ Grav.\  {\bf 40} (2008) 607
  [arXiv:0708.4231 [hep-th]].
%
\bibitem{Burgess:2013ara}
  C.~P.~Burgess,
  arXiv:1309.4133 [hep-th].
%
\bibitem{Padilla:2015aaa}
  A.~Padilla,
  arXiv:1502.05296 [hep-th].
%
\bibitem{Luty:2012ww}
  M.~A.~Luty, J.~Polchinski and R.~Rattazzi,
  JHEP {\bf 1301} (2013) 152
  [arXiv:1204.5221 [hep-th]].
%
\bibitem{Fortin:2012hn}
  J.~F.~Fortin, B.~Grinstein and A.~Stergiou,
  JHEP {\bf 1301} (2013) 184
  doi:10.1007/JHEP01(2013)184
  [arXiv:1208.3674 [hep-th]].
%
\bibitem{Dymarsky:2013pqa}
  A.~Dymarsky, Z.~Komargodski, A.~Schwimmer and S.~Theisen,
  JHEP {\bf 1510} (2015) 171
  [arXiv:1309.2921 [hep-th]].
%
\bibitem{Dymarsky:2014zja}
  A.~Dymarsky, K.~Farnsworth, Z.~Komargodski, M.~A.~Luty and V.~Prilepina,
  JHEP {\bf 1602} (2016) 099
  [arXiv:1402.6322 [hep-th]].
%
\bibitem{Gabadadze:2014rwa}
  G.~Gabadadze,
  Phys.\ Lett.\ B {\bf 739} (2014) 263
  [arXiv:1406.6701 [hep-th]];
%
  G.~Gabadadze and S.~Yu,
  arXiv:1510.07943 [hep-th].
%
\bibitem{Tseytlin:1990hn}
  A.~A.~Tseytlin,
  Phys.\ Rev.\ Lett.\  {\bf 66} (1991) 545.
%
\bibitem{Banks:2010zn}
  T.~Banks and N.~Seiberg,
  Phys.\ Rev.\ D {\bf 83} (2011) 084019
  [arXiv:1011.5120 [hep-th]].
%
\bibitem{Klinkhamer:2007pe}
  F.~R.~Klinkhamer and G.~E.~Volovik,
  Phys.\ Rev.\ D {\bf 77} (2008) 085015
  [arXiv:0711.3170 [gr-qc]].
%
\bibitem{Anderson:1971pn}
  J.~L.~Anderson and D.~Finkelstein,
  Am.\ J.\ Phys.\  {\bf 39} (1971) 901.
%
\bibitem{vanderBij:1981ym}
  J.~J.~van der Bij, H.~van Dam and Y.~J.~Ng,
  Physica {\bf 116A} (1982) 307.
%
\bibitem{Kaloper:2013zca}
  N.~Kaloper and A.~Padilla,
  Phys.\ Rev.\ Lett.\  {\bf 112} (2014) 9,  091304
  [arXiv:1309.6562 [hep-th]];
%
  Phys.\ Rev.\ D {\bf 90} (2014) 8,  084023
  [arXiv:1406.0711 [hep-th]].
%
\bibitem{Ben-Dayan:2015nva}
  I.~Ben-Dayan, R.~Richter, F.~R{\"u}hle and A.~Westphal,
  arXiv:1507.04158 [hep-th].
%
\bibitem{Nakayama:2013is}
  Y.~Nakayama,
  Phys.\ Rept.\  {\bf 569} (2015) 1
  [arXiv:1302.0884 [hep-th]].
%
\bibitem{Aurilia:1980xj}
  A.~Aurilia, H.~Nicolai and P.~K.~Townsend,
  Nucl.\ Phys.\ B {\bf 176} (1980) 509.
%
\bibitem{Duff:1980qv}
  M.~J.~Duff and P.~van Nieuwenhuizen,
  Phys.\ Lett.\ B {\bf 94} (1980) 179.
%
\bibitem{Hawking:1984hk}
  S.~W.~Hawking,
  Phys.\ Lett.\ B {\bf 134} (1984) 403.
%
\bibitem{Bousso:2000xa}
  R.~Bousso and J.~Polchinski,
  JHEP {\bf 0006} (2000) 006
  [hep-th/0004134].
%
\bibitem{Fukuyama:1983hv}
  T.~Fukuyama,
  Annals Phys.\  {\bf 157} (1984) 321.
%
\bibitem{Guendelman:1999qt}
  E.~I.~Guendelman,
  Mod.\ Phys.\ Lett.\ A {\bf 14} (1999) 1043
  [gr-qc/9901017];
%
%
  E.~I.~Guendelman, H.~Nishino and S.~Rajpoot,
  Phys.\ Lett.\ B {\bf 732} (2014) 156
  [arXiv:1403.4199 [hep-th]].
%
%
\bibitem{Guendelman:2013ke}
  E.~Guendelman, H.~Nishino and S.~Rajpoot,
  Phys.\ Rev.\ D {\bf 87} (2013) 2,  027702.
%
\bibitem{VanNieuwenhuizen:1973fi}
  P.~Van Nieuwenhuizen,
  Nucl.\ Phys.\ B {\bf 60} (1973) 478.
%
\bibitem{deRham:2010ik}
  C.~de Rham and G.~Gabadadze,
  Phys.\ Rev.\ D {\bf 82} (2010) 044020
  [arXiv:1007.0443 [hep-th]];
%
  C.~de Rham, G.~Gabadadze and A.~J.~Tolley,
  Phys.\ Rev.\ Lett.\  {\bf 106} (2011) 231101
  [arXiv:1011.1232 [hep-th]].
%
\bibitem{Hassan:2011hr}
  S.~F.~Hassan and R.~A.~Rosen,
  Phys.\ Rev.\ Lett.\  {\bf 108} (2012) 041101
  [arXiv:1106.3344 [hep-th]];
%
  JHEP {\bf 1204} (2012) 123
  [arXiv:1111.2070 [hep-th]].
%
\bibitem{Hinterbichler:2011tt}
  K.~Hinterbichler,
  Rev.\ Mod.\ Phys.\  {\bf 84} (2012) 671
  [arXiv:1105.3735 [hep-th]].
%
\bibitem{deRham:2014zqa}
  C.~de Rham,
  Living Rev.\ Rel.\  {\bf 17} (2014) 7
  [arXiv:1401.4173 [hep-th]].
%
\bibitem{Duff:1989ah}
  M.~J.~Duff,
  Phys.\ Lett.\ B {\bf 226} (1989) 36
   [Conf.\ Proc.\ C {\bf 8903131} (1989) 403].
%
\bibitem{Polchinski:2016xto}
  J.~Polchinski,
  arXiv:1601.06145 [hep-th].
%
\bibitem{ArkaniHamed:2002sp}
  N.~Arkani-Hamed, H.~Georgi and M.~D.~Schwartz,
  Annals Phys.\  {\bf 305} (2003) 96
  [hep-th/0210184].
%
\bibitem{Randall:1999ee}
  L.~Randall and R.~Sundrum,
  Phys.\ Rev.\ Lett.\  {\bf 83} (1999) 3370
  [hep-ph/9905221].
%
\bibitem{Randall:1999vf}
  L.~Randall and R.~Sundrum,
  Phys.\ Rev.\ Lett.\  {\bf 83} (1999) 4690
  [hep-th/9906064].
%
\bibitem{DeWolfe:1999cp}
  O.~DeWolfe, D.~Z.~Freedman, S.~S.~Gubser and A.~Karch,
  Phys.\ Rev.\ D {\bf 62} (2000) 046008
  [hep-th/9909134].
%
\bibitem{Kachru:2000hf}
  S.~Kachru, M.~B.~Schulz and E.~Silverstein,
  Phys.\ Rev.\ D {\bf 62} (2000) 045021
  [hep-th/0001206].
%
\bibitem{ArkaniHamed:2000eg}
  N.~Arkani-Hamed, S.~Dimopoulos, N.~Kaloper and R.~Sundrum,
  Phys.\ Lett.\ B {\bf 480} (2000) 193
  [hep-th/0001197].
\bibitem{Forste:2000ps}
  S.~F{\"o}rste, Z.~Lalak, S.~Lavignac and H.~P.~Nilles,
  Phys.\ Lett.\ B {\bf 481} (2000) 360
  [hep-th/0002164];
%
  JHEP {\bf 0009} (2000) 034
  [hep-th/0006139].
%
\bibitem{Csaki:2000wz}
  C.~Csaki, J.~Erlich, C.~Grojean and T.~J.~Hollowood,
  Nucl.\ Phys.\ B {\bf 584} (2000) 359
  [hep-th/0004133].
%
\bibitem{Kim:2000mc}
  J.~E.~Kim, B.~Kyae and H.~M.~Lee,
  Phys.\ Rev.\ Lett.\  {\bf 86} (2001) 4223
  [hep-th/0011118];
%
  Nucl.\ Phys.\ B {\bf 613} (2001) 306
  [hep-th/0101027];
%
  K.~S.~Choi, J.~E.~Kim and H.~M.~Lee,
  J.\ Korean Phys.\ Soc.\  {\bf 40} (2002) 207
  [hep-th/0201055];
%
  J.~E.~Kim and H.~M.~Lee,
  JHEP {\bf 0209} (2002) 052
  [hep-th/0207260];
%
  J.~E.~Kim,
  JHEP {\bf 0301} (2003) 042
  [hep-th/0210117].
%
\bibitem{Pospelov:2004aw}
  M.~Pospelov,
  Int.\ J.\ Mod.\ Phys.\ A {\bf 23} (2008) 881
  [hep-ph/0412280].
%
\bibitem{Antoniadis:2010ik}
  I.~Antoniadis, S.~Cotsakis and I.~Klaoudatou,
  Class.\ Quant.\ Grav.\  {\bf 27} (2010) 235018
  [arXiv:1010.6175 [gr-qc]];
%
  Fortsch.\ Phys.\  {\bf 61} (2013) 20
  [arXiv:1206.0090 [hep-th]].
%
\bibitem{Medved:2001ad}
  A.~J.~M.~Medved,
  J.\ Phys.\ G {\bf 28} (2002) 1169
  [hep-th/0109180].
%
\bibitem{Forste:2011hq}
  S.~F{\"o}rste, H.~P.~Nilles and I.~Zavala,
  JCAP {\bf 1107} (2011) 007
  [arXiv:1104.2570 [hep-th]].
%
\bibitem{Forste:2013hc}
  S.~F{\"o}rste, J.~E.~Kim and H.~M.~Lee,
  JCAP {\bf 1303} (2013) 022
  [arXiv:1301.4228 [hep-th]].
%
\bibitem{Antoniadis:2014ioa}
  I.~Antoniadis, S.~Cotsakis and I.~Klaoudatou,
  Eur.\ Phys.\ J.\ C {\bf 74} (2014) 12,  3192
  [arXiv:1406.0611 [hep-th]].
%
\end{thebibliography}
\end{document}